\documentstyle[aps,prl,epsf]{revtex}
\begin{document}
% for two column  activate the line below...
\twocolumn[\hsize\textwidth\columnwidth\hsize\csname@twocolumnfalse\endcsname
\author{M.\ Iavarone, G.\ Karapetrov, A.\ E.\ Koshelev, W.\ K.\ Kwok, G.\ W.\ Crabtree, and D.\ G.\ Hinks}
\address{Materials Science Division, Argonne National Laboratory,
Argonne, Illinois 60439\\}
%\author{}
%\address{}
\title{Two-Band Superconductivity in MgB$_2$}
\date{\today}
\tighten \maketitle
\begin{abstract}
The study of the anisotropic superconductor MgB$_2$ using a
combination of scanning tunneling microscopy and spectroscopy
reveals two distinct energy gaps at $\Delta_1$=2.3 meV and
$\Delta_2$=7.1 meV. Different spectral weights of the partial
superconducting density of states (PDOS) are a reflection of
different tunneling directions in this multi-band system. Our
experimental observations are consistent with the existence of
two-band superconductivity in the presence of interband
superconducting pair interaction and  quasiparticle scattering.
Temperature evolution of the tunneling spectra follows the BCS
scenario~\cite{Suhl} with both gaps vanishing at the bulk T$_c$.
Indeed, the study of tunneling junctions exhibiting only the small
gap (c-axis tunneling) clearly and reproducibly show that this gap
persists up to the bulk T$_c$. The data confirm the importance of
Fermi-surface sheet dependent superconductivity in MgB$_2$
proposed in the multigap model by Liu et al.~\cite{Liu} .
\end{abstract}

\vskip.2pc] \narrowtext
%\vspace{-0.4in}
%\begin{multicols}{2}

The discovery of superconductivity in MgB$_2$~\cite{Akimitsu} at
39K sparked great interest in the fundamental physics and
practical applications of this material. There has already been
rapid progress in understanding the physical properties of this
superconductor. Specific heat measurements~\cite{Bouquet,Wang}
show that MgB$_2$ is an s-wave superconductor and the presence of
the isotope effect~\cite{Bud'ko,Hinks} points towards
phonon-mediated pairing. Tunneling and photoemission spectroscopy
directly measures the superconducting energy gap and can provide
further understanding of the origin of the superconductivity in
this material. Earlier tunneling spectroscopy measurements show a
large spread in the gap values~\cite{Karapetrov,Rubio,Schmidt}
each consistent with the BCS form. More recent experiments,
including STM tunneling spectroscopy ~\cite{Giubileo},
point-contact spectroscopy~\cite{Szabo,herbert}, specific heat
measurements ~\cite{Bouquet,Wang}, and Raman
spectroscopy~\cite{Chen} point towards the existence of two
distinct gaps. This scenario has been predicted theoretically by
Liu et al.~\cite{Liu}. First principle calculations show that the
Fermi surface of MgB$_2$ consists of 2D cylindrical sheets arising
from $\sigma$ antibonding states of B $p_{xy}$ orbitals, and 3D
tubular networks arising from $\pi$ bonding and antibonding states
of B $p_z$ orbitals. In this theoretical framework~\cite{Liu} two
different energy gaps exist, the smaller one being an induced gap
associated with the 3D bands and the larger one associated with
the superconducting 2D bands. Furthermore both superconducting
gaps should vanish at the bulk critical temperature T$_c$. Due to
this highly anisotropic band structure the superconducting gaps
should be momentum-dependent reflecting the strength of the
electron-phonon coupling of the carriers in the different bands.
Up to now there has been no direct experimental evidence of the
orientation dependence of the order parameter in this material.
Moreover, the temperature dependence of the two gaps would give
further insights into the nature of superconductivity in MgB$_2$.
Scanning tunneling spectroscopy is a unique technique that allows
direct measure of the DOS near the Fermi energy with high spatial
and energy resolution. Since the tunneling current couples only to
bands with an appreciable Fermi velocity in the tunneling
direction STM can distinguish between the 2D bands with little or
no c-axis component of Fermi velocity and 3D bands with
considerable c-axis component. Thus, in the clean limit($l \gg
2\pi\xi$) the direction of the tunneling current selects which
band and which superconducting gap is probed. In this paper we
report direct evidence of orientation-dependent double-gap
structure in the quasi-particle energy spectra as determined from
tunneling spectroscopy. Our data are consistent with the
theoretical predictions of multigap superconductivity in the clean
limit. Temperature dependent tunneling shows that the two distinct
gaps vanish simultaneously near the bulk T$_c$.

%is particularly sensitive to the crystallographic orientation of
%the electrodes different directions of quasiparticle tunneling
%should probe different bands in the case of a superconductor in
%the clean limit .

Compact samples of MgB$_{2}$ were synthesized from amorphous B
powder (4N's purity) and high purity Mg. The B powder was pressed
into pellets under 6 kbar pressure. These pellets were reacted
with Mg vapor at 850 C for 2 hr in a BN container under 50 bar of
Ar. During the reaction the pellets broke up in pieces of several
mm in size. The typical critical temperature of these gold-colored
pellets is 39K.

The tunneling measurements were performed on these samples using
different surface preparation methods. One procedure consisted of
chemically etching the pellets for 50 seconds in bromine (Br 1{\%}
in pure ethanol), then rinsing in pure ethanol and drying in
N$_{2}$ gas.  Alternative procedures consisted of cleaving or
mechanical polishing of the sample in an inert atmosphere. After
these treatments the samples were mounted on the STM stage in He
exchange gas and cooled down to 4.2 K. These procedures all
yielded equivalent tunneling results. The measurements were
performed with a home-built STM described
elsewhere~\cite{Karapetrov}.

% Current-voltage characteristics
%(I-V) and conductance spectra (dI/dV vs V) were recorded at
%different locations on the scanning area. The differential
%conductance dI/dV vs V curves were recorded using a standard
%lock-in technique with a small ac modulation superimposed on a
%slowly varying bias voltage while the feedback loop was
%interrupted. The amplitude of the ac-modulation was fixed between
%0.2 mV and 0.4 mV, well below the intrinsic thermal broadening at
%4.2K.

The typical conductance spectra recorded on different grains
within the sample are shown in Figure~\ref{Figure1}. The typical
grain size is of the order of 50 nm and the spectra are remarkably
reproducible within the same grain both with respect to location
and tunneling resistance. All the spectra are normalized to the
conductance value at --20mV. They reveal a double gap structure,
flat background, and very low zero-bias conductance with very
little broadening other than thermal smearing. Two peaks are
present at V$_{1}\sim$~3 mV and V$_{2}\sim$~7.5 mV, symmetrically
for both injection and emission of electrons and they stay within
10{\%} of these values when changing between grains and/or
samples. This gap size distribution is in line with the proposed
distribution of the gap values over the Fermi surface by Choi et
al.~\cite{Choi}.

Figure~\ref{Figure2}(a) is the representative surface topography
(150 nm x 150 nm) showing several different grains. The roughness
within an individual grain is about 1nm and the grains are
separated by steps that are 1-10nm high. Each of the four grains
is characterized by a tunneling spectrum that is consistent within
a particular grain. Within this scanning area we focus on the
tunneling spectra acquired on grains 1 and 2
(Fig.~\ref{Figure2}a). The spectrum recorded in grain 2 shows a
clear two-gap structure dominated by the high-energy gap while in
grain 1 only the smaller gap is pronounced and the high-energy
peak becomes a weak satellite feature. The main characteristics of
the spectra, including the spectral weight of each peak, do not
change with the tunneling resistance from 0.1 to 2 G$\Omega $.
Having this remarkable spatial reproducibility of the tunneling
conductance we focus our attention on the origin of the different
spectra in these grains. First, we point out that both spectra are
in agreement with the multi-gap scenario proposed by Liu et
al.~\cite{Liu} where in the clean limit the impurity and surface
scattering do not significantly influence the spectra.
%The signatures of two gaps in each grain is an clear
%evidence of existence of two distinct order parameters. This
%observation is consistent with the sample being in the clean limit
%when $l \gg \xi $ and therefore the impurity and surface
%scattering do not significantly influence the spectra.
Second, we find a striking similarity between the spectrum in
grain 1 and the typical spectrum we recorded earlier on c-axis
oriented films~\cite{BobO}. In the c-axis tunneling on MgB$_2$
epitaxial films the contribution from the 3D Fermi surface is
expected to dominate the tunneling conductance. Thus we believe
that the distinct spectra attributable to grain 1 and 2 are a
reflection of different crystallographic grain orientation with
respect to the tunneling direction. This evidence of
momentum-dependent tunneling is further supported by the spatial
and temperature-dependent analysis of the spectra presented below.

To elucidate the origin of the different spectra in the two grains
we studied the spatial evolution of the spectra across the grain
boundary. In Figure~\ref{Figure2}(b,c) we show the conductance
maps at the superconducting peak values of V$_1$=+3.3 mV and
V$_2$=+7.3 mV in the region indicated by the square spanning the
two grains. In these maps the bright zones mark regions where the
tunneling spectra show higher conductance at the energy
corresponding to V$_{1}$ (Fig.~\ref{Figure2}(b)) or V$_{2}$
(Fig.~\ref{Figure2}(c)), respectively. The transition from one
type of spectra to the other evolves over a length scale on the
order of 5 nm. We associate these two different types of spectra
with vacuum tunneling into grains with two distinct orientations.
The smooth transition between the spectra across the grain
boundary on {\it the length scale of} $\xi$ is further evidence of
the sample being in the clean limit.

% begin Koshelev's stuff
The distinct tunneling conductance spectra that we observe are
consistent with the two-gap BCS model ~\cite{ManyBands} taking
into account interband impurity scattering (see, e.g.,
~\cite{SungJPCS67}).
According to this model the partial densities of states (PDOS) of two bands $%
N_{1 (2)}(\omega )$
\[
N_{1(2)}(\omega )=N_{1(2)}(0)%
%TCIMACRO{\func{Re}}%
%BeginExpansion
\mathop{\rm Re}%
%EndExpansion
\left[ \frac{u_{1(2)}}{\sqrt{u_{1(2)}^{2}-1}}\right] ,
\]
are determined by two dimensionless functions $u_{1 (2)}(\omega )$
that obey equations
\begin{eqnarray*}
u_{1}\Delta _{1} &=&\omega +i\Gamma +i\Gamma _{12}\frac{u_{2}-u_{1}}{\sqrt{u_{2}^{2}-1%
}},
\label{eq1}
\\
u_{2}\Delta _{2} &=&\omega +i\Gamma +i\Gamma _{21}\frac{u_{1}-u_{2}}{\sqrt{u_{1}^{2}-1%
}}
\label{eq2}
.
\end{eqnarray*}
Here $\Delta _{1,2}$ are the gap parameters of two bands, $\Gamma
_{12}$ and $\Gamma _{21}$ are the interband scattering rates, with
$\Gamma _{12}/\Gamma _{21}=N_{2}(0)/N_{1}(0)$. It is important to
note that strong interband scattering suppresses the
superconducting transition temperature. In particular, at $\Gamma
_{21}\ll T_{c}$ and $\Delta _{1}\ll \Delta _{2}$, $\delta
T_{c}\approx -\pi \Gamma _{21}/8$ (see, e.
g.,~\cite{GolubMazPRB97}).

%!!!Just in case, these are gap equations (probably not needed in
%the paper)!!!
%\begin{eqnarray*}
%\frac{\Delta _{1}U_{22}-\Delta
%_{2}U_{12}}{U_{11}U_{22}-U_{12}^{2}} &=&N_{1}\int_{0}^{\omega
%_{D}}d\omega
%%TCIMACRO{\func{Re}}%
%%BeginExpansion
%\mathop{\rm Re}%
%%EndExpansion
%\left[ \frac{1}{\sqrt{u_{1}^{2}-1}}\right] \tanh \frac{\omega }{2T} \\
%\frac{\Delta _{2}U_{11}-\Delta
%_{1}U_{12}}{U_{11}U_{22}-U_{12}^{2}} &=&N_{2}\int_{0}^{\omega
%_{D}}d\omega
%%TCIMACRO{\func{Re}}%
%%BeginExpansion
%\mathop{\rm Re}%
%%EndExpansion
%\left[ \frac{1}{\sqrt{u_{2}^{2}-1}}\right] \tanh \frac{\omega
%}{2T}
%\end{eqnarray*}

% end Koshelev's stuff

We use a linear combination of the two PDOS to obtain the total
DOS

\[
N_{tot} (\omega ) = \alpha N_1 (\omega ) + \beta N_2 (\omega )
\]

\noindent that we compare with the experimental tunneling
conductance spectra. The PDOS' spectral weight represented by the
parameters $\alpha $ and $\beta $ ($\alpha +\beta $=1) is
associated with the relative orientation of the crystallographic
axis of a grain with respect to the tunneling barrier. Thus in the
fits shown below the parameters $\alpha $ and $\beta $ are kept
constant for all spectra obtained on a particular grain at
different temperatures.

The linear combination of the two PDOS shown above accounts for
all major features in our tunneling spectra including the
temperature dependent data. The evolution of the tunneling spectra
in the temperature range between 4.2K and 42K are displayed in
Figure~\ref{Figure4}. The two-gap signature in the spectra of
grain 1 becomes hardly distinguishable with increasing the
temperature (Fig.~\ref{Figure4}(a)). The small gap shows a rapid
decrease in the peak height and the conductance peaks of the two
gaps merge into a broad peak at T$\approx$10K. In order to
separate the temperature evolution of each PDOS we used the
temperature dependence of the spectrum on grain 2
(Fig.~\ref{Figure4}(b)) where only the small gap is present. Each
curve was normalized to the conductance value at -20 mV and then
compared with the theoretical curve at the same temperature. The
theoretical curves were obtained by fixing the interband
scattering rates $\Gamma _{12}$, $\Gamma _{21}$ and the smearing
parameter $\Gamma$ at values optimized for the 4.2 K conductance
spectra.
%The
%theoretical curves have been obtained using a common set of
%parameters (gap values $\Delta_1$(T) and $\Delta_2$(T), interband
%scattering parameters $\Gamma_{12}$ and $\Gamma_{21}$, and the
%pairbreaking parameter $\Gamma$) with fixed values of the relative
%contributions from the two bands (parameters $\alpha$ and $\beta$)
%for each grain.
In all the fits we put $\Gamma _{12}$= 0.73 $\Gamma _{21}$ since
their ratio is proportional to the ratio of the density of states
at the Fermi level of the 3D and the 2D
bands~\cite{SungJPCS67,Schopohl}. The theoretical curves reproduce
very well the shape inside the gap but not the peak heights. By
adjusting the parameters $\Gamma _{12}$ and $\Gamma$ we could
reproduce very well the peak heights but the zero bias value would
be higher than the experimental one. Fits to the two set of
parameters have been used to get an estimate of the error bars for
the two gap values: $\Delta _{1}$= 2.3$\pm $0.2 mV and $\Delta
_{2}$= 7.1$\pm $0.2. A very good fit of both peaks' height and
zero bias value can be obtained only for interband scattering
values that are very large ($\sim $2.5 mV) and values of the 3D
gap is extremely small. According to~\cite{GolubMazPRB97} such a
large interband scattering value should lower the critical
temperature by $ \sim$9 K which contradicts our experimental
findings. The set of temperature dependent tunneling spectra
clearly shows that both superconducting gaps exist up to 37K.

Next, we turn our attention to the spectral weights of the two
PDOS to the total DOS. The particular choice of the two grains in
Fig.~\ref{Figure2} was based on the peculiarity of the conductance
spectra of these crystallites. They closely represent the two
extreme limits of partial contribution by each PDOS that we have
observed so far on a large number of pellets and thin films. The
theoretical fits obtained by our model in these two limits
(Fig.~\ref{Figure4}a,b) show that the contribution of the 3D PDOS
varies from nearly 100{\%} (Fig.~\ref{Figure4}b and ~\cite{BobO})
down to 87{\%} (Fig.~\ref{Figure4}a). Therefore it appears that
momentum averaged STM tunneling conductance is dominated by the 3D
band in this momentum dependent two-band tunneling process. This
could be due to the specific geometry of the tunneling cone
emanating from the STM tip and its projection on the highly
anisotropic Fermi surface of the MgB$_{2}$ crystallite. Obviously,
the relative weights of the PDOS are temperature independent and
the parameters $\alpha $ and $\beta $ remain fixed when the
temperature dependent fits of the tunneling spectra are obtained.

The values of $\Delta _{1}$(T) and $\Delta _{2}$(T) are extracted
from the theoretical curves and the results are reported in
Figure~\ref{Figure4}(c). The high-energy gap follows a BCS
behavior. The smaller gap remains constant up to 20K and follows
the BCS predictions with a critical temperature far beyond the one
expected for such a small gap (T$_{c} \approx $14K) with a ratio
2$\Delta$/kT$_{c}\approx$1.4 similar to earlier findings from
point spectroscopy~\cite{herbert}. Such behavior clearly
demonstrates that the gap in the 3D band is mainly induced by the
interband pairing interaction rather than produced by pairing
interaction inside this band, in agreement with theoretical
expectations \cite{Liu,Choi}. This experimental evidence directly
excludes the possibility that the grains or their surfaces are
associated with different stoichiometry. The high-energy gap
follows a BCS behavior with a ratio $2\Delta_2(0)/kT_{c}\approx
4.3$, significantly larger than the BCS value of 3.52. There are
two possible sources for this increase. Usually such enhancement
is attributed to strong coupling effect. However, this ratio also
increases due to the presence of additional superconducting bands
even in the weak coupling limit. In the case of
$\Delta_1\ll\Delta_2$ one can derive a simple formula from the
two-band BCS theory
\[
\frac{2\Delta _{2}}{T_{c}}\approx 3.52\left( 1+\frac{N_{1}\Delta _{1}^{2}}{%
N_{2}\Delta _{2}^{2}}\ln \frac{\Delta _{2}}{\Delta _{1}}\right).
\]
Using $N_2=0.73 N_1$ from \cite{Liu} and experimental values of
the gaps, we estimate from this formula $2\Delta _{2}/T_{c}\approx
4.1$. Therefore, we conclude that the main part of the $2\Delta
/T_{c}$ increase originates from the multiband nature of the
superconductivity in this compound.

Finally, we would like to point out that in the case of two-band
superconductivity in the dirty limit ($l\ll\xi)$ the conductance
spectra are very broad and similar to those reported
earlier~\cite{Karapetrov}. These spectra are mostly obtained on
pellets without any surface treatment. Impurities near the surface
may provide sufficiently strong interband scattering to smear out
the two-gap features in this case.

In conclusion, the existence of two gaps has been observed by
tunneling spectroscopy in compact pellets of MgB$_{2}$. Tunneling
spectra show different ratios between the two gap heights on
different grains within the sample. This result can be interpreted
in terms of different tunneling directions with respect to the
crystallographic orientation of the grain, therefore supporting
the two-gap scenario~\cite{Liu,Choi}. Different tunneling
direction should probe the two bands with different weights.
Moreover, the two gaps both vanish at a temperature close to the
bulk value. The observed temperature independence of the small gap
up to T$ \approx $ 20 K implies that no transition temperature
corresponding to the BCS ratio 2$\Delta _{1}$=3.52 kT$_{c}$ is
present. This suggests that the coupling between the two bands is
moderately strong and that the superconductivity is dominant in
the 2D sheets.

This work was supported by US DOE Basic Energy Science - Material
Science under contract No. W-31-109-ENG-38.

\begin{figure}
\epsfxsize=3.1in \epsffile{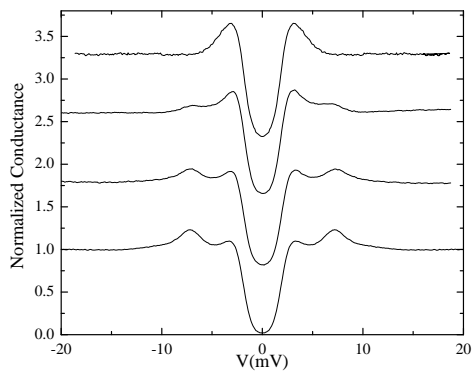} \caption{Tunneling
conductance spectra recorded on different grains at T=4.2K. The
tunneling resistance is 0.2 G$\Omega $. The spectra have been
normalized to the conductance value at high voltage.}
\label{Figure1}
\end{figure}

\begin{figure}
\epsfxsize=3.1in \epsffile{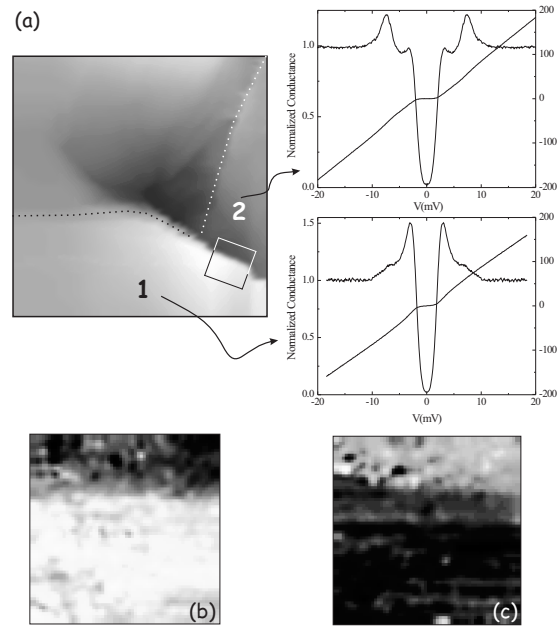} \caption{(a) Topographic
image of the scanning area 150 nm x 150 nm acquired in the
constant current mode (sample bias V= -20 mV, I= 200 pA). The
conductance spectra and the current-voltage characteristics on two
different grains are shown on the right. The tunneling resistance
is 0.2 G$\Omega $. (b) and (c) Conductance maps of the scanning
area 30 nm x 30 nm outlined in image (a) by the square across the
grain boundary. The conductance maps have been recorded at
V$_1$=3.3 mV (b) and V$_2$=7.3 mV (c) respectively.}
\label{Figure2}
\end{figure}

%\begin{figure}
%\epsfxsize=3.1in \epsffile{mgb2f3.eps} \caption{Tunneling spectra
%recorded on grains 1 and 2 together with fitted theoretical
%conductance curves. The experimental curves have been normalized
%to the high voltage conductance and the tunneling resistance is
%0.2 G$\Omega $. Solid lines: theoretical curves, using equations
%(1) and (2), with the parameters (a) $\Delta _{1}$=2 mV, $\Delta
%_{2}$= 6.9 mV, $\Gamma _{12}$=0.3 mV, $\Gamma _{21}$=0.5mV,
%$\Gamma $=0.11mV, $\alpha $=0.8, $\beta $=0.2 (b) $\Delta _{1}$=2
%mV, $\Delta _{2}$= 7 mV, $\Gamma _{12}$=0.5 mV, $\Gamma
%_{21}$=0.7mV, $\Gamma $=0.11mV, $\alpha $=0.97, $\beta $=0.03.
%$\alpha $ and $\beta $ are the spectral weights of the two bands.}
%\label{Figure3}
%\end{figure}

\begin{figure}
\epsfxsize=4.1in \epsffile{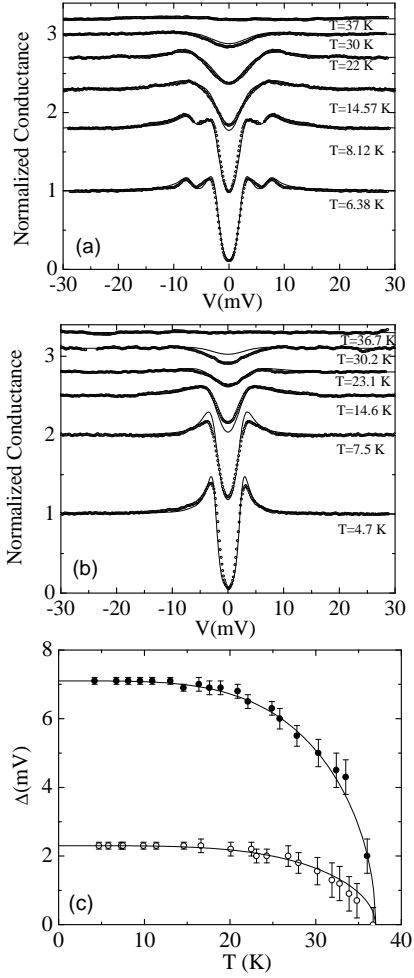} \caption{Temperature
evolution of the two tunneling spectra together with the
theoretical curves. For all the curves $\Gamma _{12}$=0.1 mV,
$\Gamma _{21}$=0.14mV, $\Gamma $=0.11mV (values optimized to the
experimental curves at 4.2 K). In (a) the curves are reproduced
using $\alpha $= 0.87, $\beta $=0.13. In (b) $\alpha $= 1, $\beta
$=0. In (c) the gap values are extracted from the theoretical
curves and are plotted as a function of the temperature together
with the BCS $\Delta $(T).} \label{Figure4}
\end{figure}

\end{document}